\documentclass[italian,english]{article}
\usepackage[T1]{fontenc}
\usepackage[latin9]{inputenc}
\usepackage{color}
\usepackage{amssymb}
\usepackage{esint}
\usepackage{babel}
\begin{document}

\title{\textbf{Cosmology and gravitational waves in the Nordstr\"{o}m-Vlasov
system, a laboratory for Dark Energy}}

\author{\textbf{Christian Corda}}

\maketitle
\begin{center}
Institute for Theoretical Physics and Advanced Mathematics Einstein-Galilei,
Via Santa Gonda 14, 59100 Prato, Italy 
\par\end{center}

\begin{center}
and 
\par\end{center}

\begin{center}
Inter-University Centre Engineering of life and Environment, LIUM
University, Via Lugano 2, 6500 Bellinzona, Switzerland
\par\end{center}

\begin{center}
\textit{E-mail addresses:} \textcolor{blue}{cordac.galilei@gmail.com} 
\par\end{center}
\begin{abstract}
We discuss a cosmological solution of the system which was originally
introduced by Calogero and is today popularly known as ``Nordstr\"{o}m-Vlasov
system''. 

Although the model is unphysical, its cosmological solution results
interesting for the same reasons for which the Nordstr\"{o}m-Vlasov
system was originally introduced in the framework of galactic dynamics.
In fact, it represents a theoretical laboratory where one can rigorously
study some problems, like the importance of the gravitational waves
in the dynamics, which at the present time are not well understood
within the physical model of the Einstein\textendash{}Vlasov system.

As the cosmology of the Nordstr\"{o}m-Vlasov system is founded on
a scalar field, a better understanding of the system is important
also in the framework of the Dark Energy problem. In fact, various
attempts to achieve Dark Energy by using scalar fields are present
in the literature.

In the solution an analytical expression for the time dependence of
the cosmological evolution of the Nordstr\"{o}m's scalar field is
also released. This analytical expression is unknown in the literature
of the Nordstr\"{o}m-Vlasov system.

Based on their importance, the propagation of gravitational waves
in the Nordstr\"{o}m-Vlasov system and their effects on test masses
are also discussed.\end{abstract}
\begin{quotation}
\textbf{Keywords: Nordström-Vlasov system; gravitational waves, Dark
Energy.}
\end{quotation}

\section{Introduction}

Historically, the relativistic scalar theory of gravitation introduced
in 1912-13 by the Finnish physicist Gunnar Nordstr\"{o}m \cite{key-1}
has been the first metric theory of gravity. In fact, it was derived
three years before Einstein's general relativity \cite{key-2}. Although
Nordstr\"{o}m's scalar theory of gravity is today ruled out by observations,
it results an interesting example of non-minimal coupling of matter
to gravity and shares with general relativity the unique property
to embody the strong equivalence principle \cite{key-3}. It is also
considered an useful toy model to test analytical \cite{key-4} or
numerical \cite{key-5} analyses on the two-body problem. 

The interesting idea to couple Nordstr\"{o}m's scalar theory of gravity
with the Vlasov's kinetic model for a collisionless ensemble of particles
interacting through the gravitational force is due to Calogero, who
analyzed the ``Nordstrom-Vlasov system'' in the framework of galactic
dynamics \cite{key-6}. Calogero and collaborators further discussed
the Nordström-Vlasov system and its solutions in refs. \cite{key-7}-\cite{key-12}.

In this paper we discuss a cosmological solution of the Nordström-Vlasov
system. The cosmological framework is interesting for the same reasons
for which the system was originally introduced in the framework of
galactic dynamics \cite{key-6}. It is considered an interesting theoretical
laboratory where one can rigorously study problems which at the present
time are not well understood within the physical model of the Einstein\textendash{}Vlasov
system \cite{key-6}. An example is the importance of the gravitational
waves in the dynamics \cite{key-6}. 

We recall that the cosmology of the Nordstr\"{o}m-Vlasov system is
founded on a scalar field. Hence, a better understanding of the system
is important also in the framework of the Dark Energy problem as various
attempts to achieve Dark Energy by using scalar fields are present
in the literature see \cite{key-19}-\cite{key-23} ad references
within.

The solution shows also the cosmological evolution of Nordstr\"{o}m's
scalar field which is unknown in the literature of the Nordstr\"{o}m-Vlasov
system.

Based on their importance, the propagation of gravitational waves
in the Nordstr\"{o}m-Vlasov system and their effects on test masses
are also discussed in the last Section of this paper.

\section{The Nordström-Vlasov system}

Let us clarify an important issue. One has to distinguish between
the standard kinetic theory, which is based on the Boltzmann equation,
and the Vlasov's framework and its applications, i.e. the electromagnetic
Vlasov\textendash{}Poisson system, the Vlasov\textendash{}Poisson
non-relativistic system and the relativistic Nordström-Vlasov and
Einstein-Vlasov systems. In a kinetic description the time evolution
of the system is determined by the interactions between the particles,
which depend on the physical situation \cite{key-13,key-14}. For
instance, the driving mechanism for the time evolution of a neutral
gas is the collision between particles (the Boltzmann equation) \cite{key-13,key-14}.
For a plasma the interaction is through the electromagnetic field
produced by the charges (the Vlasov\textendash{}Maxwell system), and
in astrophysics the interaction between collisionless particles is
gravitational (the Vlasov\textendash{}Poisson system and the Nordström-Vlasov
and Einstein-Vlasov systems) \cite{key-13,key-14}. Thus, an important
difference between the standard kinetic theory and the Vlasov's framework
is that in the first case there are collisions between particles while
in the Vlasov's framework particles interact electromagnetically or
gravitationally only and they are \emph{collisionless} \cite{key-13,key-14}. 

If one wants to satisfy the condition demanding that the particles
make up an ensemble with no collisions in the spacetime, the particle
density must be a solution of the Vlasov equation \cite{key-14,key-15} 

\begin{equation}
\partial_{t}f+\frac{p^{a}}{p^{0}}\partial_{x^{a}}f-\Gamma_{\mu\nu}^{a}\frac{p^{\mu}p^{\nu}}{p^{0}}\partial_{p^{a}}f=0.\label{eq: Vlasov}
\end{equation}
In this paper we work with $8\pi G=1$, $c=1$ and $\hbar=1,$ while
the sign conventions for the line element, which generate the sign
conventions for the Riemann/Ricci tensors, are $(-,+,+,+).$ Latin
indices run from 1 to 3, and Greek from 0 to 3 and $x^{0}=t$. Here
$\Gamma_{\mu\nu}^{\alpha}$ represent the usual connections, $f$
is the particle density and $p^{0}$ is determined by $p^{a}$($a=1,2,3$)
according to the relation \cite{key-14,key-15} 
\begin{equation}
g_{\mu\nu}p^{\mu}p^{\nu}=-1,\label{eq: mass-shell}
\end{equation}
which expresses the condition that the four momentum $p^{\mu}$ lies
on the mass shell of the metric. 

We recall that, in general, the Vlasov-Poisson system is \cite{key-14,key-15}

\begin{equation}
\begin{array}{c}
\partial_{t}f+v\cdot\bigtriangledown_{x}f-\bigtriangledown_{x}U\cdot\bigtriangledown_{v}f=0\\
\\
\bigtriangleup U=4\pi\rho\\
\\
\rho(t,x)=\int dvf(t,x,v),
\end{array}\label{eq: VP}
\end{equation}

where $t$ denotes the time and $x$ and $v$ the position and the
velocity of the particles. The function $U=U(t,x)$ is the average
Newtonian potential generated by the particles. This system represents
the non-relativistic kinetic model for an ensemble of particles with
no collisions, which interacts through the gravitational forces that
they generate collectively \cite{key-14,key-15}. Thus, one can use
such a system to describe the motion of galaxies within the Universe,
thought of as pointlike particles, when the relativistic effects are
negligible \cite{key-14}. In this approach, the function $f(t,x,v)$
in the Vlasov-Poisson system (\ref{eq: VP}) is non-negative and gives
the density on phase space of the galaxies within the Universe \cite{key-14}.
If the relativistic effects become not negligible, the motion of the
particles has to be described by the Einstein\textendash{}Vlasov system
\cite{key-6}. In this relativistic system, the dynamics of the matter
is still described by a non-negative scalar function $f$, but the
unknown of the field equations is now the metric of the space-time
\cite{key-6}. 

The Einstein\textendash{}Vlasov system is much more complicated than
the Vlasov\textendash{}Poisson system \cite{key-6}. The greatest
difficulties are due to the hyperbolic and highly non-linear character
of the Einstein field equations, to the equivalence of all the coordinate
systems in general relativity and to the issue that in this theory
of gravitation the space-time is itself part of the solution of the
equations \cite{key-6}. In fact, obtaining global solutions of the
Einstein field equations coupled to any kind of matter is extremely
difficult \cite{key-16,key-17}. 

For these reasons Calogero \cite{key-6} proposed a simplified model,
the Nordström-Vlasov system, in which the dynamics of the matter is
still described by the Vlasov equation but where the gravitational
forces between the particles are now supposed to be mediated by Nordstr\"{o}m's
scalar theory of gravity \cite{key-1}. In such a theory the gravitational
forces are mediated by a scalar field $\phi$ and the effect of such
forces is to induce a curvature in the space-time \cite{key-1,key-6}.
Moreover, the scalar field modifies the otherwise flat metric only
by a rescaling \cite{key-1,key-6}. Hence, the spacetime of the model
is given by the conformally flat metric \cite{key-1,key-6}

\begin{equation}
ds^{2}=A^{2}(\phi)\left(dt^{2}-dz^{2}-dx^{2}-dy^{2}\right).\label{eq: metrica puramente scalare}
\end{equation}

If one requests the property of \emph{scale invariance} the function
$A$ results set as

\begin{equation}
A(\phi)=\exp(\phi)\label{eq: set}
\end{equation}
 (except a constant which is fixed equal to one), see \cite{key-6}
for details.

\section{Cosmological solution}

By assuming a time dependence of the Nordstr\"{o}m's scalar field
$\phi\equiv\phi(t)$, the line element (\ref{eq: metrica puramente scalare})
results analogous to the cosmological Friedmann-Robertson-Walker (FRW)
line element of the standard homogeneous, isotropic and flat Universe
\cite{key-16,key-17}

\begin{equation}
ds^{2}=A^{2}\left[\phi(t)\right]\left(dt^{2}-dz^{2}-dx^{2}-dy^{2}\right),\label{eq: metrica FRW}
\end{equation}

and the function $A\left[\phi(t)\right]$ results to be the scale
factor. 

By using the classical transformation from conformal time to synchronous
time \cite{key-14,key-16,key-17} 

\begin{equation}
dt\rightarrow\frac{dt}{A\left[\phi(t)\right]}\label{eq: trasformazione temporale}
\end{equation}

the line element (\ref{eq: metrica puramente scalare}), in spherical
coordinates, becomes 
\begin{equation}
ds^{2}=dt^{2}-A^{2}\left[\phi(t)\right](dr^{2}+r^{2}(d\theta^{2}+\sin^{2}\theta d\varphi^{2})).\label{eq: metrica puramente scalare piena 2}
\end{equation}

The metric tensor has the form \cite{key-14,key-16,key-17} 

\begin{equation}
g_{\mu\nu}=\left(\begin{array}{ccc}
1 &  & 0\\
\\
0 &  & \gamma_{mn}
\end{array}\right),\label{eq: tensore metrico}
\end{equation}

where $\gamma_{mn}=A^{2}\left[\phi(t)\right].$

Let us define \cite{key-14} $\chi_{mn}\equiv\frac{\partial}{\partial t}\gamma_{mn.}$
The Einstein field equations in the synchronous frame are \cite{key-14,key-16,key-17}
\begin{equation}
R_{0}^{0}=-\frac{1}{2}\frac{\partial}{\partial t}\chi_{a}^{a}-\frac{1}{4}\chi_{a}^{b}\chi_{b}^{a}=(T_{0}^{0}-\frac{1}{2}T)\label{eq: sincrona 1}
\end{equation}

\begin{equation}
R_{a}^{0}=\frac{1}{2}(\chi_{a;b}^{b}-\chi_{b;a}^{a})=T_{a}^{0}\label{eq: sincrona 2}
\end{equation}

\begin{equation}
R_{a}^{b}=-P_{a}^{b}-\frac{1}{2\sqrt{\gamma}}\frac{\partial}{\partial t}(\sqrt{\gamma}\chi_{a}^{b})=T_{a}^{b}-\frac{1}{2}\delta_{a}^{b}T,\label{eq: sincrona 3}
\end{equation}

where $P_{a}^{b}$ is the Ricci tensor in $3$ dimensions \cite{key-14,key-16,key-17}
.

On the other hand, the Einstein field equations in the Vlasov's framework
are \cite{key-14,key-15}

\begin{equation}
G_{\mu\nu}=\frac{2}{\sqrt{-g}}\int f(t,x,p)p_{\mu}p_{\nu}\delta(p^{2}+m^{2})d^{4}p,\label{eq: G}
\end{equation}

where $m$ is the mass of a particle (galaxy).

Following \cite{key-14} we can split the function $f(t,x,p)$ into
a couple of equations for $f_{+}(t,x,p)$ and $f_{-}(t,x,p)$ which
are constructed by reducing $f(t,x,p)$ respectively on the \textquotedblright{}upper\textquotedblright{}
half and on the \textquotedblright{}lower\textquotedblright{} half
of the mass shell. Eq. (\ref{eq: Vlasov}) becomes \cite{key-14}

\begin{equation}
\partial_{t}f_{\pm}=-\frac{1}{p_{\pm}^{0}}\left(\gamma^{mn}p_{n}\frac{\partial}{\partial x_{m}}-\frac{1}{2}\frac{\partial\gamma^{nr}}{\partial x_{m}}p_{n}p_{r}\frac{\partial}{\partial p_{m}}\right)f_{\pm}.\label{eq: Vlasov 2}
\end{equation}

Eq. (\ref{eq: Vlasov 2}) can be interpreted in Hamiltonian terms
\cite{key-14}

\begin{equation}
p_{\pm}^{0}\partial_{t}f_{\pm}=\{H,f_{\pm}\},\label{eq: interpretazione hamiltoniana}
\end{equation}

where the Hamiltonian function is \cite{key-14}

\begin{equation}
H\equiv\frac{1}{2}\gamma^{mn}p_{m}p_{n}.\label{eq: hamiltoniana}
\end{equation}

One can calculate the components of energy-momentum tensor $T_{\mu\nu}$
in the approximation which considers galaxies like massless particles
\cite{key-14} ($m=0$ in Eq. (\ref{eq: G}))

\begin{equation}
T_{00}=\frac{1}{A^{3}\left[\phi(t)\right]r^{2}\sin\theta}\int\frac{f_{+}+f_{-}}{A\left[\phi(t)\right]}\sqrt{\frac{p_{1}^{2}+p_{2}^{2}}{r^{2}}+\frac{p_{3}^{2}}{r^{2}\sin\theta}}d^{3}p\label{eq: zero-zero}
\end{equation}

\begin{equation}
T_{mn}=\frac{1}{A^{3}\left[\phi(t)\right]r^{2}\sin\theta}\int A(\phi)\frac{(f_{+}+f_{-})}{\sqrt{\frac{p_{1}^{2}+p_{2}^{2}}{r^{2}}+\frac{p_{3}^{2}}{r^{2}\sin\theta}}}p_{m}p_{n}d^{3}p\label{eq: m-n}
\end{equation}

\begin{equation}
T_{0m}=\frac{1}{A^{3}\left[\phi(t)\right]r^{2}\sin^{2}\theta}\int(f_{+}-f_{-})p_{m}d^{3}p.\label{eq: zero-m}
\end{equation}

The Einstein field equations (\ref{eq: sincrona 1}), (\ref{eq: sincrona 2})
and (\ref{eq: sincrona 3}) give two independent dynamic equations
which can be written down in terms of the scale factor $A(\phi)$:

\begin{equation}
\dot{A}^{2}\left[\phi(t)\right]=-1+\frac{1}{3A\left[\phi(t)\right]}\int(f_{+}(s)+f_{-}(s))\frac{s}{A\left[\phi(t)\right]}d^{3}s\label{eq: s1}
\end{equation}

\begin{equation}
\ddot{A}\left[\phi(t)\right]=-\frac{2}{A\left[\phi(t)\right]}-2\frac{\dot{A}^{2}\left[\phi(t)\right]}{A\left[\phi(t)\right]}+\frac{1}{A^{2}\left[\phi(t)\right]}\int(f_{+}(s)+f_{-}(s))d^{3}s,\label{eq: s2}
\end{equation}

where \cite{key-14} 
\begin{equation}
s\equiv p_{1}^{2}+\frac{p_{2}^{2}}{r^{2}}+\frac{p_{3}^{2}}{r^{2}\sin^{2}\theta}.\label{eq: s}
\end{equation}

By introducing the dimensionless variables $\underline{r}$ and $\underline{t}$
\cite{key-14} we put 

\begin{equation}
\begin{array}{c}
A\left[\phi(t)\right]=A(\phi_{0})\underline{r}\\
\\
t=A(\phi_{0})\underline{t}\\
\\
\dot{\underline{r}}=\frac{d\underline{r}}{d\underline{t}}\\
\\
j=\frac{1}{3}A^{2}(\phi_{0})\rho_{0},
\end{array}\label{eq: new variables}
\end{equation}

where $A(\phi_{0})\equiv A\left[\phi(t_{0})\right]$ is the present
value of the scale factor of the Universe being $\phi_{0}\equiv\phi(t_{0})$
the present value of the Nordstr\"{o}m's scalar field.

Eq. (\ref{eq: s1}) becomes \cite{key-14} 
\begin{equation}
\begin{array}{c}
\dot{\underline{r}}^{2}=-1+\frac{j}{\underline{r}^{2}}\\
\\
\underline{r}_{0}=1.
\end{array}\label{eq: system}
\end{equation}

The solution of the system (\ref{eq: system}) is \cite{key-14}

\begin{equation}
\underline{r}(\underline{t})=\sqrt{j-(\underline{t}-\sqrt{j-1})^{2}}\label{eq: solution}
\end{equation}

if $j\geq1.$ Notice that expression (\ref{eq: solution}) becomes
imaginary for certain values of the parameter $\underline{t}$. Such
a parameter represents the cosmic time normalized by the present value
of the scale factor $A(\phi_{0})$, see eq. (\ref{eq: new variables}).
By choosing the origin of the cosmic time at the present era of the
cosmological evolution, i.e. $t_{0}=0,$ expression (\ref{eq: solution})
is \emph{always} real if $j\geq1.$ 

Returning to the ($t,$ $A\left[\phi(t)\right]$) variables we get:

\begin{equation}
A\left[\phi(t)\right]=A(\phi_{0})\sqrt{\frac{A^{2}(\phi_{0})\rho_{0}}{3}-(\frac{t}{A(\phi_{0})}-\sqrt{\frac{A^{2}(\phi_{0})\rho_{0}}{3}}-1)^{2}}.\label{eq: soluzione}
\end{equation}

The today's observed Hubble radius and the today's observed density
of the Universe are \foreignlanguage{italian}{respectively \cite{key-18}
$A(\phi_{0})\gtrsim10^{28}{\normalcolor cm}$ and $\rho_{0}\approx10^{-57}{\normalcolor cm^{-2}}.$
Therefore $j\approx1.$}

If one inserts these data in Eq. (\ref{eq: soluzione}) a singularity
at a time

$t_{s}\approx-10^{10}{\normalcolor seconds}$ and a value for the
today's theoretical Hubble constant $H_{0}=\frac{\dot{A}(\phi_{0})}{A(\phi_{0})}\approx10^{-29}{\normalcolor cm^{-1}}$
are obtained.

Although the discussed model is unphysical, we note that, even under
the assumption to neglect the baryon mass of the galaxies, the results
look reasonable as they are of the same order of magnitude of the
standard cosmological model \foreignlanguage{italian}{\cite{key-18}}. 

By using eq. (\ref{eq: set}), eq. (\ref{eq: soluzione}) permits
to obtain the cosmological evolution of the Nordstr\"{o}m's scalar
field 
\begin{equation}
\phi(t)=\phi_{0}+\frac{1}{2}\ln\left[\frac{\rho_{0}\exp(2\phi_{0})}{3}-\left(\frac{t}{\exp(\phi_{0})}-\sqrt{\frac{\rho_{0}\exp(2\phi_{0})}{3}}-1\right)^{2}\right].\label{eq: soluzione campo scalare}
\end{equation}

We emphasize that this is the first time that an analytical expression
for the time dependence of the Nordstr\"{o}m's scalar field is released.
In fact, the expression (\ref{eq: soluzione campo scalare}) is unknown
in the literature of the Nordström-Vlasov system.

\section{Propagation of gravitational waves and motion of test masses in the
Nordstr\"{o}m-Vlasov system}

Let us linearize the line element (\ref{eq: metrica puramente scalare}).
Putting $\phi=\phi_{*}+\frac{h}{2}$ with $h\ll1$ and using (\ref{eq: set}),
eq. (\ref{eq: metrica puramente scalare}) becomes 

\begin{equation}
\begin{array}{c}
ds^{2}=\exp\left(2\phi_{*}+h\right)\left(dt^{2}-dz^{2}-dx^{2}-dy^{2}\right)=\\
\\
=\exp(2\phi_{*})\exp(h)\left(dt^{2}-dz^{2}-dx^{2}-dy^{2}\right).
\end{array}\label{eq: linearizza 1}
\end{equation}

We can also normalize the line element (\ref{eq: linearizza 1}) to
$\exp(2\phi_{*})$ without loss of generality. Hence, to first order
in $h,$ 
\begin{equation}
ds^{2}=\left(1+h\right)\left(dt^{2}-dz^{2}-dx^{2}-dy^{2}\right).\label{eq: linearizza 2}
\end{equation}

As we are interested to the propagation of gravitational waves, we
consider the linearized theory on vacuum by setting equal to zero
the right hand side of the Einstein field equation (\ref{eq: G}).
Following \cite{key-16}, to first order in $h$, we call $\widetilde{R}_{\mu\nu\rho\sigma}$,
$\widetilde{R}_{\mu\nu}$ and $\widetilde{R}$ the linearized quantity
which correspond to $R_{\mu\nu\rho\sigma}$, $R_{\mu\nu}$ and $R$.
By inserting the line element (\ref{eq: linearizza 2}) in the linearized
field equations 
\begin{equation}
\widetilde{R}_{\mu\nu}-\frac{\widetilde{R}}{2}\eta_{\mu\nu}=0,\label{eq: linearizzate}
\end{equation}

we obtain only one independent equation, the d'alembertian

\begin{equation}
\square h=0.\label{eq: dalembertiana}
\end{equation}

The solution of eq. (\ref{eq: dalembertiana}) is the plane wave 
\begin{equation}
h=a(\overrightarrow{k})\exp(ik^{\alpha}x_{\alpha})+c.c.\label{eq: sol S}
\end{equation}

which represents a scalar massless mode. Considering a wave propagating
in the positive $z$ direction, the line element (\ref{eq: linearizza 2})
reads

\begin{equation}
ds^{2}=\left[1+h(t-z)\right]\left(dt^{2}-dz^{2}-dx^{2}-dy^{2}\right).\label{eq: metrica finale}
\end{equation}

Now, let us discuss the effect of these gravitational waves on two
test masses $m_{1}$ and $m_{2}$. We will use the coordinate system
in which the space-time is locally flat and the distance between any
two points is given simply by the difference in their coordinates
in the sense of Newtonian physics \cite{key-16}. This frame is the
proper reference frame of a local observer, located for example in
the position O of the first test mass $m_{1}$. In this frame gravitational
waves manifest them-self by exerting tidal forces on the two test
masses \cite{key-16}. A detailed analysis of the frame of the local
observer is given in Sect. 13.6 of \cite{key-16}. Here, only the
more important features of this frame are recalled \cite{key-16}:

the time coordinate $x_{0}$ is the proper time of the observer O;

spatial axes are centered in O;

in the special case of zero acceleration and zero rotation the spatial
coordinates $x_{j}$ are the proper distances along the axes and the
frame of the local observer reduces to a local Lorentz frame. In this
case the line element reads \cite{key-16}

\begin{equation}
ds^{2}=-(dx^{0})^{2}+\delta_{ij}dx^{i}dx^{j}+O(|x^{j}|^{2})dx^{\alpha}dx^{\beta}.\label{eq: metrica local lorentz}
\end{equation}

The effect of the gravitational wave on test masses is described by
the equation \cite{key-16}

\begin{equation}
\ddot{x^{i}}=-\widetilde{R}_{0k0}^{i}x^{k},\label{eq: deviazione geodetiche}
\end{equation}
which is the equation for geodesic deviation in this frame.

Thus, to study the effect of the gravitational wave on test masses,
$\widetilde{R}_{0k0}^{i}$ has to be computed in the proper reference
frame of the local observer. But, because the linearized Riemann tensor
$\widetilde{R}_{\mu\nu\rho\sigma}$ is invariant under gauge transformations
\cite{key-16}, it can be directly computed from eq. (\ref{eq: metrica finale}). 

From \cite{key-16} one gets

\begin{equation}
\widetilde{R}_{\mu\nu\rho\sigma}=\frac{1}{2}\{\partial_{\mu}\partial_{\beta}h_{\alpha\nu}+\partial_{\nu}\partial_{\alpha}h_{\mu\beta}-\partial_{\alpha}\partial_{\beta}h_{\mu\nu}-\partial_{\mu}\partial_{\nu}h_{\alpha\beta}\},\label{eq: riemann lineare}
\end{equation}

that, in the case eq. (\ref{eq: metrica finale}), as a sole polarization
is present, begins

\begin{equation}
\widetilde{R}_{0\gamma0}^{\alpha}=\frac{1}{2}\{\partial^{\alpha}\partial_{0}h\eta_{0\gamma}+\partial_{0}\partial_{\gamma}h\delta_{0}^{\alpha}-\partial^{\alpha}\partial_{\gamma}h\eta_{00}-\partial_{0}\partial_{0}h\delta_{\gamma}^{\alpha}\};\label{eq: riemann lin scalare}
\end{equation}

the different elements are (only the non zero ones will be written
down explicitly):

\begin{equation}
\partial^{\alpha}\partial_{0}h\eta_{0\gamma}=\left\{ \begin{array}{ccc}
\partial_{t}^{2}h & for & \alpha=\gamma=0\\
\\
-\partial_{z}\partial_{t}h & for & \alpha=3;\gamma=0
\end{array}\right\} \label{eq: calcoli}
\end{equation}

\begin{equation}
\partial_{0}\partial_{\gamma}h\delta_{0}^{\alpha}=\left\{ \begin{array}{ccc}
\partial_{t}^{2}h & for & \alpha=\gamma=0\\
\\
\partial_{t}\partial_{z}h & for & \alpha=0;\gamma=3
\end{array}\right\} \label{eq: calcoli2}
\end{equation}

\begin{equation}
-\partial^{\alpha}\partial_{\gamma}h\eta_{00}=\partial^{\alpha}\partial_{\gamma}h=\left\{ \begin{array}{ccc}
-\partial_{t}^{2}h & for & \alpha=\gamma=0\\
\\
\partial_{z}^{2}h & for & \alpha=\gamma=3\\
\\
-\partial_{t}\partial_{z}h & for & \alpha=0;\gamma=3\\
\\
\partial_{z}\partial_{t}h & for & \alpha=3;\gamma=0
\end{array}\right\} \label{eq: calcoli3}
\end{equation}

\begin{equation}
-\partial_{0}\partial_{0}h\delta_{\gamma}^{\alpha}=\begin{array}{ccc}
-\partial_{z}^{2}h & for & \alpha=\gamma\end{array}.\label{eq: calcoli4}
\end{equation}

Now, putting these results in eq. (\ref{eq: riemann lin scalare})
one gets

\begin{equation}
\begin{array}{c}
\widetilde{R}_{010}^{1}=-\frac{1}{2}\ddot{h}\\
\\
\widetilde{R}_{010}^{2}=-\frac{1}{2}\ddot{h}\\
\\
\widetilde{R}_{030}^{3}=\frac{1}{2}\square h.
\end{array}\label{eq: componenti riemann}
\end{equation}

The third of eqs. (\ref{eq: componenti riemann}) together with eq.
(\ref{eq: dalembertiana}) gives $\widetilde{R}_{030}^{3}=0.$ Hence,
we have shown that gravitational waves in the Nordstr\"{o}m's theory
of gravity are transverse to the direction of propagation. In fact,
using eq. (\ref{eq: deviazione geodetiche}) one gets

\begin{equation}
\ddot{x}=\frac{1}{2}\ddot{h}x,\label{eq: accelerazione mareale lungo x}
\end{equation}

\begin{equation}
\ddot{y}=\frac{1}{2}\ddot{h}y.\label{eq: accelerazione mareale lungo y}
\end{equation}

Equations (\ref{eq: accelerazione mareale lungo x}) and (\ref{eq: accelerazione mareale lungo y})
give the tidal acceleration of the test mass $m_{2}$ caused by the
gravitational wave respectly in the $x$ direction and in the $y$
direction. 

Equivalently, one can say that there is a gravitational potential
\cite{key-16}

\begin{equation}
V(\overrightarrow{r},t)=-\frac{1}{4}\ddot{h}(t)[x^{2}+y^{2}],\label{eq:potenziale in gauge Lorentziana}
\end{equation}

which generates the tidal forces, and that the motion of the test
mass is governed by the Newtonian equation \cite{key-16}

\begin{equation}
\ddot{\overrightarrow{r}}=-\bigtriangledown V.\label{eq: Newtoniana}
\end{equation}

The solution of eqs. (\ref{eq: accelerazione mareale lungo x}) and
(\ref{eq: accelerazione mareale lungo y}) can be found by using the
perturbation method \cite{key-16}. To first order in $h$, the displacement
of the test mass due by the gravitational wave are given by

\begin{equation}
\delta x(t)=\frac{1}{2}x_{0}h(t)\label{eq: spostamento lungo x}
\end{equation}

and

\begin{equation}
\delta y(t)=\frac{1}{2}y_{0}h(t),\label{eq: spostamento lungo y}
\end{equation}

where $x_{0}$ and $y_{0}$ are the initial (unperturbed) coordinates
of the test mass $m_{2}$.

\subsection*{Conclusion remarks}

In this paper a cosmological solution of the Nordstr\"{o}m-Vlasov
system has been discussed by releasing an analytical expression for
the time dependence of the Nordstr\"{o}m's scalar field which is unknown
in the literature. Although the model has to be considered unphysical,
this cosmological solution results interesting for the same reasons
for which the Nordstr\"{o}m-Vlasov system was originally introduced
in the framework of galactic dynamics in \cite{key-6}. In fact, it
represents a theoretical laboratory where one can rigorously study
some problems, like the importance of the gravitational waves in the
dynamics, which at the present time are not well understood within
the physical model of the Einstein\textendash{}Vlasov system. 

Based on their importance, the propagation of gravitational waves
in the Nordstr\"{o}m-Vlasov system and their effects on test masses
have also been discussed in the last Section of this paper.

The cosmology of the Nordstr\"{o}m-Vlasov system is founded on a scalar
field. Thus, a better understanding of the system is important also
in the framework of the Dark Energy problem. On this issue, we recall
that various attempts to achieve Dark Energy by using scalar fields
are present in the literature \cite{key-19}-\cite{key-23} .

\end{document}